\documentclass[10pt]{iopart}
\usepackage{epsfig}
\usepackage{subfigure}

\begin{document}

\title[Computational studies of light in straight and curved optical 
	fibres]{Computational studies of light acceptance and propagation 
	in straight and curved optical fibres} 
\author{C P Achenbach and J H Cobb}
\address{University of Oxford, Sub-department of Particle Physics, 
	Denys Wilkinson Bld., Keble Rd., Oxford, OX1 3RH, UK}
\ead{p.achenbach@physics.ox.ac.uk}

\begin{abstract}
    A Monte Carlo simulation has been performed to track light rays in
    cylindrical fibres by ray optics. The trapping efficiencies for skew
    and meridional rays in active fibres and distributions of
    characteristic quantities for all trapped light rays have been
    calculated. The simulation provides new results for curved fibres,
    where the analytical expressions are too complex to be solved. The
    light losses due to sharp bending of fibres are presented as a
    function of the ratio of curvature to fibre radius and bending
    angle. It is shown that a radius of curvature to fibre radius ratio of
    greater than 65 results in a loss of less than 10\% with the loss
    occuring in the initial stage of the bend (at bending angles $\Phi
    \sim \pi/8$\,rad).
\end{abstract}

\submitto{\JOA}

\section{Introduction}

Active optical fibres are becoming more and more important in the
field of detection and measurement of ionising radiation and
particles. Light is generated inside the fibre either through
interaction with the incident radiation (scintillating fibres) or
through absorption of primary light (wavelength-shifting
fibres). Plastic fibres with large core diameters, i.e.\ where the
wavelength of the light being transmitted is much smaller than the
fibre diameter, are commercially available and readily fabricated,
have good timing properties and allow a multitude of different
geometrical designs. The low costs of plastic materials make it
possible for many experiments to use such fibres in large quantities,
particularly in highly segmented tracking detectors and sampling
calorimeters (see reference~\cite{Leutz1995} for a review of plastic
fibres in high energy physics). Although for many years fibres have
been the subject of extensive studies, only fragmentary calculations
of trapping efficiencies and light losses in curved fibres have been
performed for multi-mode fibres. We have therefore performed a full
simulation of photons propagating in simple circularly curved fibres
in order to quantify the losses caused by the bending and to establish
the dependence of these losses on the angle of the bend. We have also
briefly investigated the time dispersion in fibres. For our
calculations the most common type of fibre in particle physics is
assumed. This standard fibre is specified by a polystyrene core of
refractive index $n_{\it core}=$ 1.6 and a thin polymethylmethacrylate
(PMMA) cladding of refractive index $n_{\it clad}=$ 1.49, where the
indices are given at a wavelength of 590\,nm. Another common cladding
material is fluorinated polymethacrylate which has a slightly lower
index of reflection of $n_{\it clad}= 1.42$. Typical diameters are in
the range of 0.5 -- 1.5\,mm.

The treatment of small diameter optical fibres involves
electromagnetic theory applied to dielectric waveguides, which was
first achieved by Snitzer~\cite{Snitzer1961} and
Kapany~\cite{Kapany1963}. Although this approach provides insight into
the phenomenon of total internal reflection and eventually leads to
results for the field distributions and electromagnetic radiation for
curved fibres, it is advantageous to use ray optics for applications
to large diameter fibres where the waveguide analysis is an
unnecessary complication. In ray optics a light ray may be
characterised by its path along the fibre. The path of a {\em
meridional ray} is confined to a single plane, all other modes of
propagation are known as {\em skew rays}. In general, the projection
of a meridional ray on a plane perpendicular to the fibre axis is a
straight line, whereas the projection of a skew ray changes its
orientation with every reflection. In the special case of a
cylindrical fibre all meridional rays pass through the fibre axis. The
optics of meridional rays in fibres was developed in the
1950s~\cite{Kapany1957} and can be found in numerous textbooks, for
example in references~\cite{Kapany1967,Allan1973}.

This paper is organised as follows: Section~2 describes the analytical
expressions of trapping efficiencies for skew and meridional rays in
active, i.e.\ light generating, fibres. The analytical description of
skew rays is too complex to be solved for sharply curved fibres and
the necessity of a simulation becomes evident. In Section~3 a
simulation code is outlined that tracks light rays in cylindrical
fibres governed by a set of geometrical rules derived from the laws of
optics. Section~4 presents the results of the simulations. These
include distributions of the characteristic properties which describe
light rays in straight and curved fibres, where special emphasis is
placed on light losses due to the sharp bending of fibres. Light
dispersion is briefly reviewed in the light of the results of the
simulation. The last section provides a short summary.

\section{Trapping of Photons}

When using scintillating or wavelength-shifting fibres in charged
particle detectors the trapped light as a fraction of the intensity of
the emitted light is important in determining the light yield of the
application. All rays which are totally internally reflected within
the cylinder of the fibre are considered as trapped. It is very well
known that the critical angle for internal reflection at the sides of
the fibre is the limiting factor (see for example~\cite{Potter1961}
and references therein). For very low light intensities as encountered
in many particle detectors the photon representation is more
appropriate to use than a description by light rays. In such
applications single photon counting is often necessary.

The geometrical path of any rays in optical fibres, including skew
rays, was first analysed in a series of papers by
Potter~\cite{Potter1961} and Kapany~\cite{Kapany1961}. The treatment
of angular dependencies in our paper is based on that. The angle
$\gamma$ is defined as the angle of the projection of the light ray in
a plane perpendicular to the axis of the fibre with respect to the
normal at the point of reflection. One may describe $\gamma$ as a
measure of the ``skewness'' of a particular ray, since meridional rays
have this angle equal to zero. The polar angle, $\theta^\prime$, is
defined as the angle of the light ray in a plane containing the fibre
axis and the point of reflection with respect to the normal at the
point of reflection. It can be shown that the angle of incidence at
the walls of the cylinder, $\alpha$, is given by $\cos{\alpha}=
\cos{\theta^\prime}\, \cos{\gamma}$. The values of the two orthogonal
angles $\theta^\prime$ and $\gamma$ will be preserved independently
for a particular photon at every reflection along its path.

In general for any ray to be internally reflected, the inequality
$\sin{\alpha} \geq \sin{\theta^\prime_{\it crit}} = n_{\it
clad}/n_{\it core}$ must be fulfilled, where the critical angle,
$\theta^\prime_{\it crit}$, is given by the index of refraction of the
fibre core, $n_{\it core}$, and that of the cladding, $n_{\it
clad}$. In the meridional approximation the above equations lead to
the well known critical angle condition for the polar angle,
$\theta^\prime \ge \theta^\prime_{\it crit}$, which describes an
acceptance cone of semi-angle, $\theta\ [= \pi/2 - \theta^\prime]$,
with respect to the fibre axis. Thus, in this approximation all light
within the forward cone will be trapped and undergo multiple total
internal reflections to emerge at the end of the fibre.

\begin{figure}[htbp]
  \begin{center}
    \subfigure[]{
	\epsfig{width= 0.47 \textwidth, file= 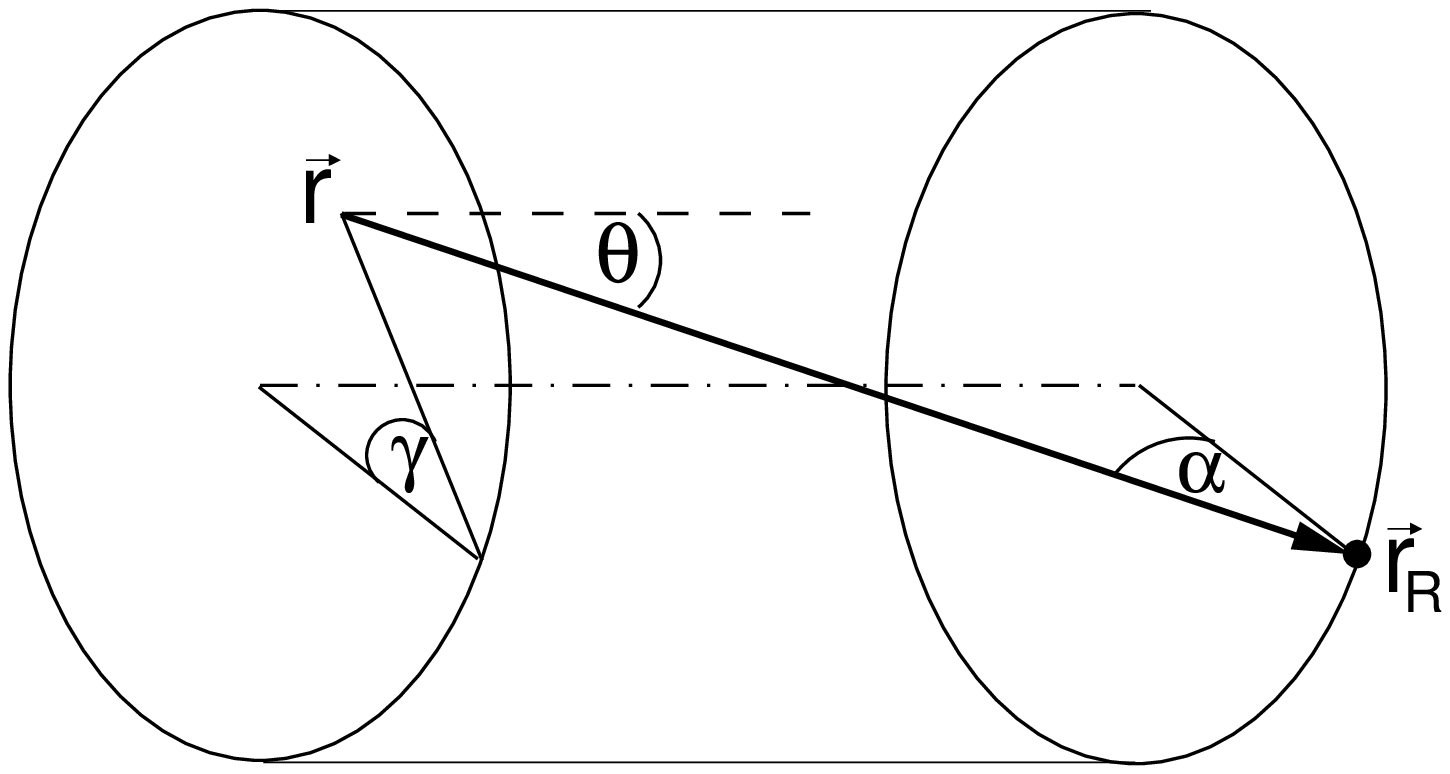} }
    \subfigure[]{
	\epsfig{width= 0.47 \textwidth, file= 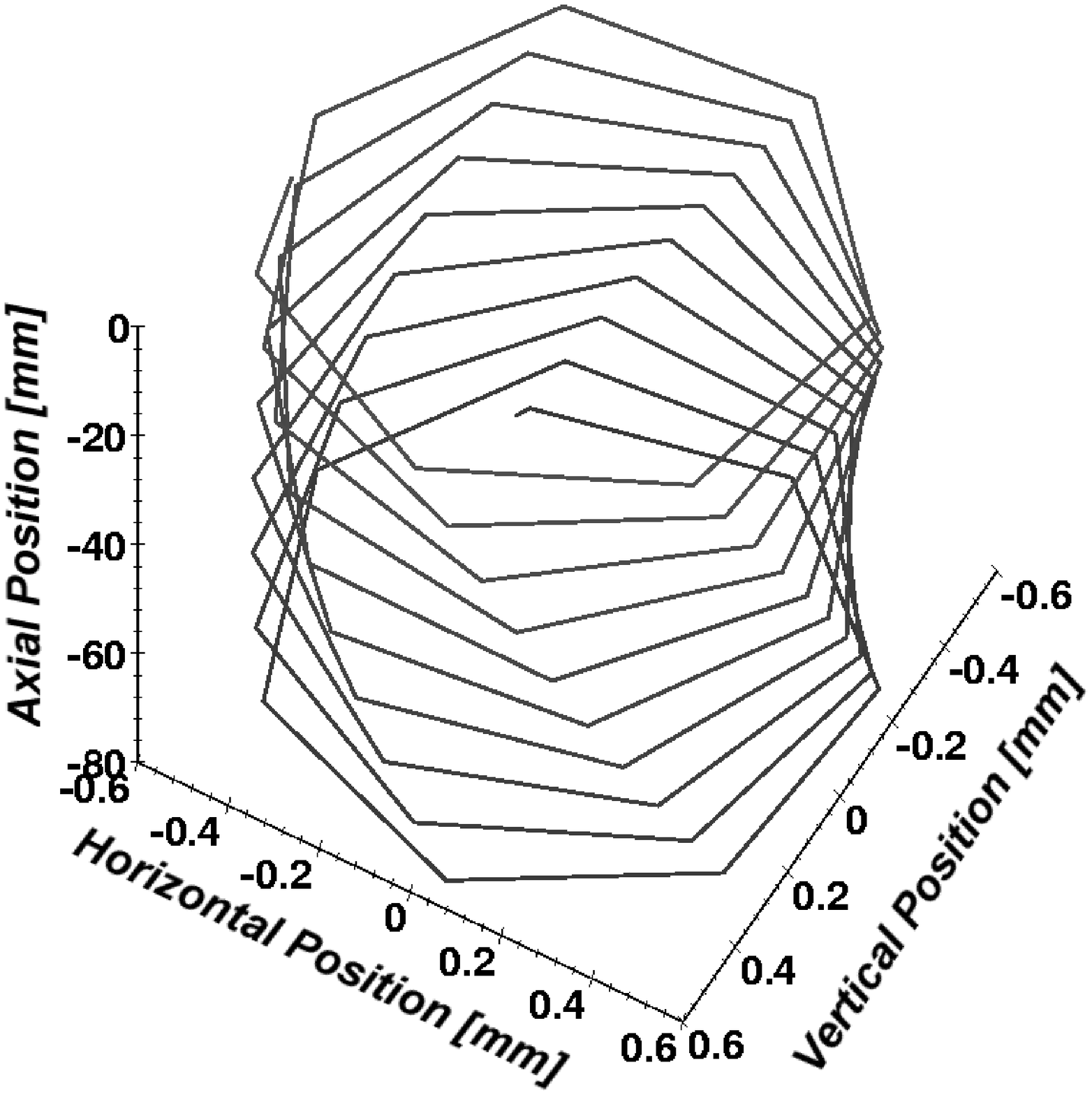} } 
    \caption{ (a) Definition of the angles for a skew ray. The 
	axial angle of the ray is $\theta$. The angle between the 
	projection of the ray in a plane perpendicular to the axis 
	of the fibre with respect to the normal at the point of 
	reflection, $\gamma$, defines the ray's ``skewness''. The 
	angle of incidence at the point of reflection is $\alpha$. 
	(b) The helical path of a skew ray.}
    \label{fig:description}
  \end{center}
\end{figure}

For the further discussion in this paper it is convenient to use the
axial angle, $\theta$, as given by the supplement of $\theta^\prime$,
and the skew angle, $\gamma$, to characterise any light ray in terms
of its orientation, see figure~\ref{fig:description}.

The flux transmitted by a fibre is determined by an integration over
the angular distribution of the light emitted within the acceptance
domain, i.e.\ the phase space of possible propagation modes. Using an
expression given by Potter~\cite{Potter1963} and setting the
transmission function which parameterises the light attenuation to
unity, the light flux can be written as follows:
\begin{equation}
\eqalign{
  F & =  F_m + F_s\\
    & =  4 \rho^2 \int_{\theta= 0}^{\theta_{\it crit}} 
	\int_{\gamma= 0}^{\pi/2} \int_{\phi= 0}^{\pi/2} 
	I(\theta,\phi)\, \cos^2{\gamma}\, d\gamma\, d\Omega\, +\\
    & 4 \rho^2 \int_{\theta= \theta_{\it crit}}^{\pi/2} 
	\int_{\gamma= \overline{\gamma}(\theta)}^{\pi/2} 
        \int_{\phi= 0}^{\pi/2}
	I(\theta,\phi)\, \cos^2{\gamma}\, d\gamma\, d\Omega,
}
\end{equation}
where $\Omega$ is the element of solid angle,
$\overline{\gamma}(\theta)$ refers to the maximum axial angle allowed
by the critical angle condition, $\rho$ is the radius of a cylindrical
fibre and $I(\theta,\phi)$ is the angular distribution of the emitted
light in the fibre core. The two terms, $F_m$ and $F_s$, refer to
either the meridional or skew cases, respectively. The lower limit of
the integral for $F_s$ is $\overline{\gamma}=
\arccos{(\sin{\theta_{\it crit}}/\sin{\theta})}$.

The trapping efficiency for forward propagating photons,
$\Omega^{1/2}$, may be defined as the fraction of totally internally
reflected photons. The formal expression for the trapping efficiency,
including skew rays, is derived by dividing the transmitted flux by
the total flux through the cross-section of the fibre core, $F_0$.
For isotropic emission of fluorescence light the total flux equals $4
\pi^2 \rho^2 I_0$. Then, the first term of equation~(1) gives the
trapping efficiency in the meridional approximation,
\begin{equation}
  \Omega^{1/2}_m = F_m/F_0 = \frac{1}{2} (1 - 
	\cos{\theta_{\it crit}}),
  \label{eq:omega_m}
\end{equation}
where all photons are considered to be trapped if $\theta \le
\theta_{\it crit}$, independent of their actual skew angles. The
latter formula yields a trapping efficiency of $\Omega^{1/2}_m=$
3.44\% for standard plastic fibres with $n_{\it core}=$ 1.6 and
$n_{\it clad}=$ 1.49.

The integration of the second term of equation~(1) gives the
contributions of all skew rays to the trapping efficiency. Integrating
by parts, one gets
\begin{equation}
\fl  \Omega^{1/2}_s = \frac{1}{2} \cos{\theta_{\it crit}} -
  \frac{\cos^2{\theta_{\it crit}} \sin{\theta_{\it crit}}}{2\pi}
  \int_0^1 \frac{dt}{\sqrt{(1-t)\,t}\, \left(1-t \cos^2{\theta_{\it
  crit}} \right)},
\end{equation}
with $t= \cos^2{\theta}/\cos^2{\theta_{\it crit}}$. Complex
integration leads to the result:
\begin{equation}
  \Omega^{1/2}_s = \frac{1}{2} (1 - \cos{\theta_{\it crit}})
  \cos{\theta_{\it crit}}.
  \label{eq:omega_s}
\end{equation}
This integral evaluates to 3.20\% for standard plastic fibres. 
The total initial trapping efficiency is then: 
\begin{equation}
  \Omega^{1/2} = \frac{1}{2} (1 - \cos^2{\theta_{\it crit}}),
\end{equation}
which is 6.64\% for standard plastic fibres, i.e.\ approximately twice
the trapping efficiency in the meridional approximation. Nevertheless,
for long fibres the effective trapping efficiency is closer to
$\Omega^{1/2}_m$ than to $\Omega^{1/2}$ since skew rays have a much
longer optical path length and therefore get attenuated more quickly
see Section~4 for a quantitative analysis.

\section{Description of the Photon Tracking Code}

The simulation code is written in Fortran. Light rays are generally
represented as lines and determined by two points, $\vec{r}$ and
$\vec{r}^{\,\prime}$. The points of incidence of rays with the
fibre-cladding boundary are determined by solving the appropriate
systems of algebraic equations. In the case of a straight fibre the
geometrical representation of a straight cylinder is used resulting in
the quadratic equation
\begin{equation} 
  (x + (x^\prime - x)\times m)^2 + (y +
  (y^\prime - y)\times m)^2 - \rho^2 = 0,
\end{equation}
where $\rho$ is the fibre radius and the fibre axis is along the
$z$-direction. The positive solution for the parameter $m$ defines the
point of incidence, $\vec{r}_R$, on the cylinder wall. In the case of
a fibre curved in a circular path, the cylinder equation is
generalised by the torus equation
\begin{equation}
  \eqalign{ \left(R - \sqrt{ (x + (x^\prime - x)\times m + R)^2 
	+ (z + (z^\prime - z)\times m)^2 } \right)^2 + \\
  	 (y + (y^\prime - y)\times m)^2 - \rho^2 = 0,}
\end{equation}
where the fibre is bent in the $xz$-plane with a radius of curvature
$R_{\it curve}$. The roots of this fourth degree polynomial are
calculated using Laguerre's method~\cite{Recipes1992}. It requires
complex arithmetic and an estimate for the root to be found. The
initial estimate is given by the intersection point of the light ray
and a straight cylinder that has been rotated and translated to the
previous reflection point. The smallest positive, real solution for
$m$ is then used to determine the reflection point, $\vec{r}_R$.

In both cases the angle of incidence, $\alpha$, is given by
$\cos{\alpha} = \vec{r}_{in} \cdot \vec{n}$, where $\vec{n}$ denotes
the unit vector normal to the fibre-cladding boundary at the point of
reflection and $\vec{r}_{in}= (\vec{r}-\vec{r}_R)/|\vec{r}-\vec{r}_R|$
is the unit incident propagation vector. The unit propagation vector
after reflection, $\vec{r}_{\it out}$, is then calculated by mirroring
$\vec{r}_{in}$ with respect to the normal vector: $\vec{r}_{\it out} =
\vec{r}_{in} - 2 \vec{n} \cos{\alpha}$.

\section{Results of the Photon Tracking Code}

Figure~\ref{fig:description} shows the passage of a skew ray along a
straight fibre. The light ray has been generated off-axis with an
axial angle of $\theta= 0.42$ and would not be trapped if it were
meridional. The figure illustrates the preservation of the skew angle,
$\gamma$, during the propagation of skew rays.

\subsection{Trapping Efficiency and Acceptance Domain}

\begin{figure}[htbp]
  \begin{center}
    \epsfig{width= 0.5 \textwidth, file= 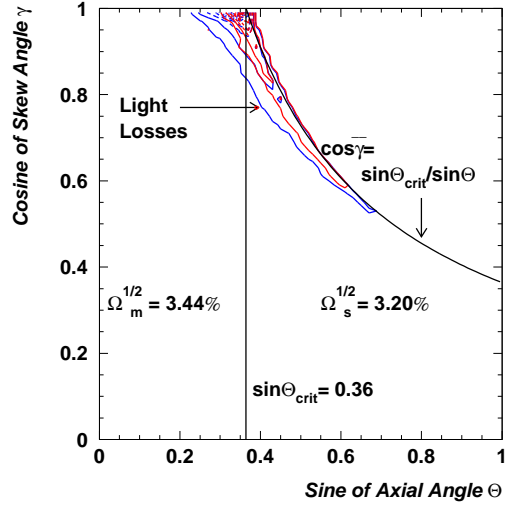}
    \caption{ The angular phase space domain for trapped photons 
	in a fibre. To the left of the dividing line at 
	$\sin{\theta_{\it crit}}$ all skew angles are accepted. To the 
	right of the line the maximum skew angle permitted by 
	the critical angle condition is given by $\cos{\overline{\gamma}}= 
	\sin{\theta_{\it crit} / \sin{\theta}}$. The trapping efficiencies 
	are evaluated by integration over the two regions. Contours of 
	phase space distributions for photons refracted out of sharply 
	curved fibres with radii of curvature $R_{\it curve}=$ 2 and 
	5\,cm are shown.}
    \label{fig:phasespace}
  \end{center}
\end{figure}

Figure~\ref{fig:phasespace} shows the total acceptance domain and its
splitting into the meridional and skew regions in the meridional ray
approximation. The figure gives the values for the two trapping
efficiencies which can be determined by integrating over the two
angular regions. The integrals are identical in value to
formulae~\ref{eq:omega_m} and~\ref{eq:omega_s}, when the photons are
generated randomly on the cross-section of the fibre with an isotropic
angular distribution in the forward direction.

A skew ray can be totally internally reflected at larger angles
$\theta$ than meridional rays and the relationship between the minimum
permitted skew angle, $\overline{\gamma}$, at a given axial angle,
$\theta$, is determined by the critical angle condition:
$\cos{\overline{\gamma}}= \sin{\theta_{\it crit}} /
\sin{\theta}$. Inside this region the phase space density is not
constant but increases with $\cos{\gamma}^2$ and $\sin{\theta}$.

\begin{figure}[htbp]
  \begin{center}
    \subfigure[]{\label{fig:trap-r}
      \epsfig{width= 0.47 \textwidth, file= 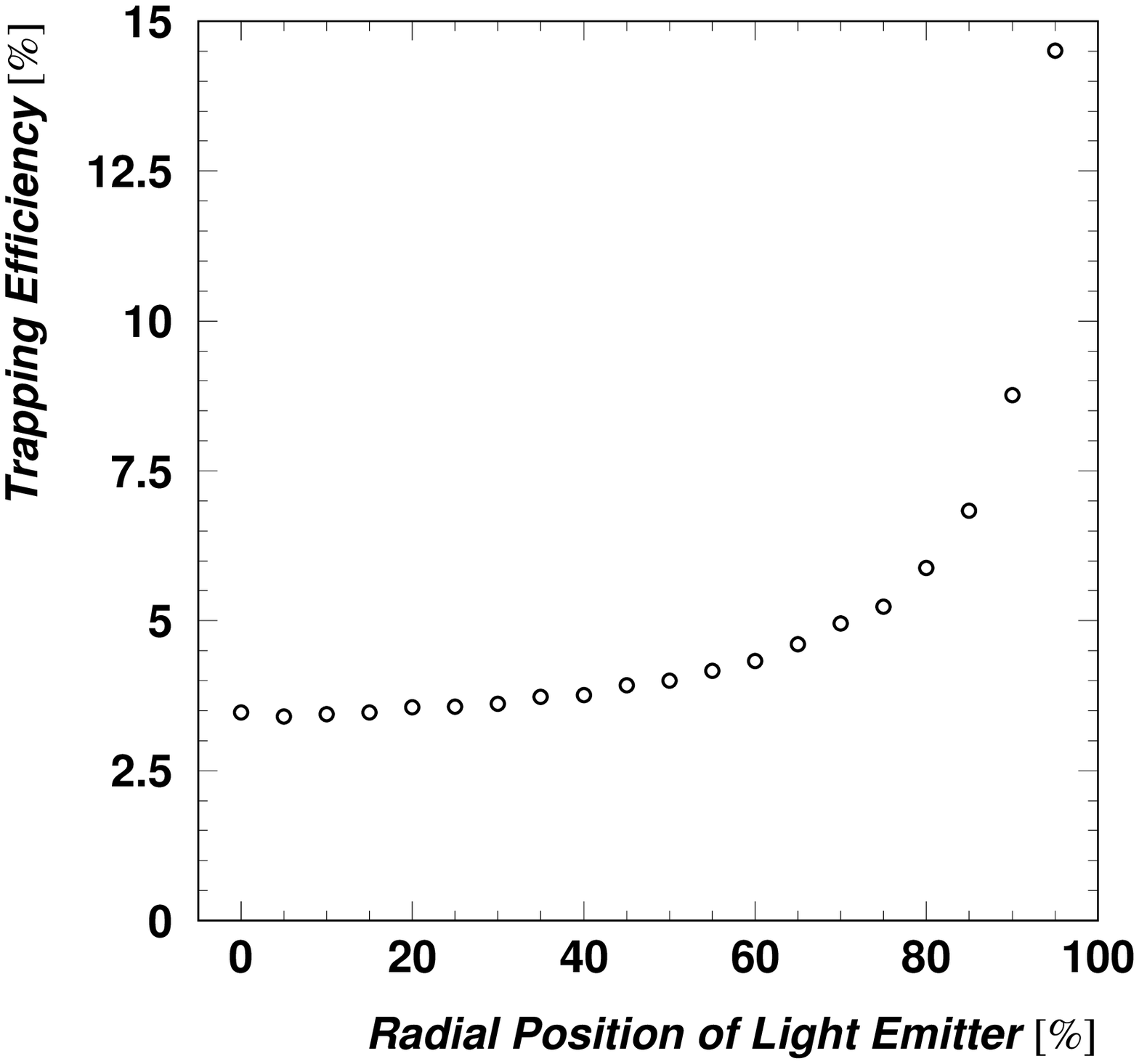} }
    \subfigure[]{\label{fig:trap-theta}
      \epsfig{width= 0.47 \textwidth, file= 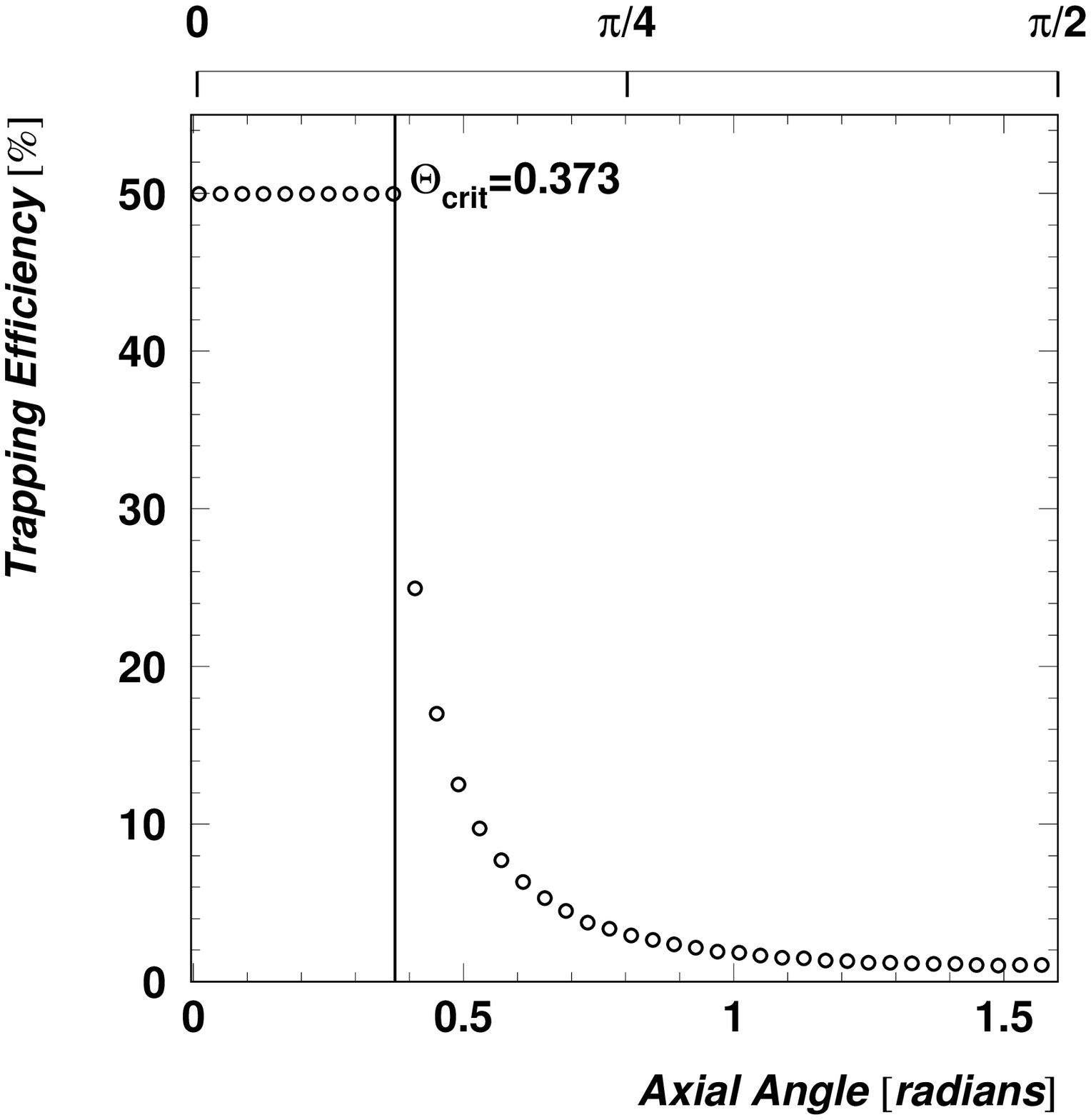} }
    \caption{ Trapping efficiency for photons propagating in 
	the forward direction as a function of radial position, 
	$\hat{\rho}$, of the light emitter in the fibre core (a)
	and of the axial angle (b). The maximum axial angle 
	allowed by the critical angle condition in the meridional 
	approximation is indicated.}
  \end{center}
\end{figure}

It is obvious from the critical angle condition that a photon emitted
close to the cladding has a higher probability to be trapped than when
emitted close to the centre of the fibre. Figure~\ref{fig:trap-r}
shows the trapping efficiency as a function of the radial position,
$\hat{\rho}$, of the light emitter in the fibre core. The trapping
efficiency is almost independent of the radial position for
$\hat{\rho} < 0.8$ and the meridional approximation, exactly valid
only at $\hat{\rho} = 0$, is a good estimate. At $\hat{\rho} \ge 0.9$
the approximation significantly underestimates the trapping
efficiency. This fact has been discussed before, e.g.\
in~\cite{Johnson1994}. Figure~\ref{fig:trap-theta} shows the the
trapping efficiency as a function of the axial angle. All photons with
axial angles below $\theta_{\it crit}$ are trapped in the fibre,
whereas photons with larger angles are trapped only if their skew
angle exceeds the minimum permitted skew angle. It can be seen that
the trapping efficiency falls off very steeply with the axial angle.

\subsection{Light Attenuation}

A fibre can conveniently be characterised by its attenuation length
over which the signal amplitude is attenuated to 1$/e$ of its original
value. However, light attenuation has many sources, among them
self-ab\-sorp\-tion, optical non-uni\-formities, reflection losses and
absorption by impurities.

Restricting the analysis to the two main sources of loss, the
transmission through a fibre can be represented for any given axial
angle by $T= \exp\left[- P(\theta) L_F/\lambda\right]\, \times q^{\eta
L_F}$, where the first term describes light losses due to bulk
absorption and scattering, and the second term describes light losses
due to imperfect reflections which can be caused by a rough surface or
variations in the refractive indices.
A comparison of some of our own measurements to determine the
attenuation length of plastic fibres with other available data
indicates that a reasonable value for the bulk absorption length is
$\lambda \sim 3$\,m. Most published data suggest a deviation of the
reflection coefficient, which parameterises the internal reflectivity,
from unity between $5 \times 10^{-5}$ and $6.5 \times 10^{-5}$
\cite{Ambrosio1991}. Only for very small diameter fibres ($\sim
10\,\mu$m) are the resulting scattering lengths of the same order as
the absorption lengths. Because of the large radii of the fibres
discussed reflection losses are not relevant for the transmission
function. A reasonable value of $q= 0.9999$ is used in the simulation
to account for all losses proportional to the number of
reflections. 

Internal reflections being less than total give rise to so-called
``leaky'' or non-guided modes, where part of the electromagnetic
energy is radiated away at the reflection points. They populate a
region defined by axial angles above the critical angle and skew
angles slightly larger than the ones for totally internally reflected
photons. These modes are taken into account by using the well known
Fresnel reflection formulas, where unpolarised light is assumed and
the reflection coefficients for the two planes of polarisation are
averaged. However, it is obvious that non-guided modes are lost
quickly.

The absorption and emission processes in fibres are spread out over a
wide band of wavelengths and the attenuation is known to be wavelength
dependent. For simplicity only monochromatic light is assumed in the
simulation and highly wavelength-dependent effects like Rayleigh
scattering are not included explicitly. Light rays are tracked in the
fibre core only and no tracking takes place in the surrounding
cladding. In long fibres cladding modes will eventually be lost, but
for lengths $< 1$\,m they can contribute to the transmission function
and will lead to a dependence of the attenuation length on the
distance from the excitation source.

\begin{figure}[htbp]
  \begin{center}
      \epsfig{width= 0.5 \textwidth, file= 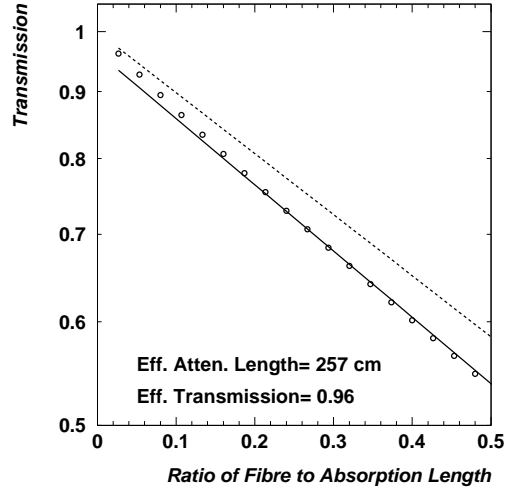}
      \caption{ Simulated transmission function for a straight fibre 
	with a bulk light absorption length $\lambda=$ 3\,m and a 
	reflection coefficient $q=$ 0.9999. The transmission as a 
	function of the ratio of fibre to absorption length, 
	$L_F/\Lambda_m$, is calculated from the optical
	path length distribution. A simple exponential fit results in 
	an effective attenuation length of $\Lambda_{\it eff}=$ 2.4\,m. 
	The dashed line shows the transmission function in the 
	meridional approximation with $\Lambda_m=$ 2.8\,m.}
      \label{fig:absorption}
  \end{center}
\end{figure}

A question of practical importance for the estimation of the light
output of a particular fibre application is its transmission
function. In the meridional approximation and substituting
$\exp(-\ln{q})$ by $\exp(1-q)$ the attenuation length can be written
as
\begin{equation}
  \Lambda_m = -\sec{\theta_{\it crit}} L_F\left[ 1/\lambda +
  (1-q)\sin{\theta_{\it crit}}/2\rho \right],
\end{equation}
which evaluates to $\Lambda_m= 2.8$\,m for the given attenuation
parameters. The correct transmission function can be found by
integrating over the normalised path length distribution (which will
be discussed in the following section):
\begin{equation}
  T = \frac{1}{N} \int_{P=0}^{\infty} dN/dP\, e^{-PL/\lambda}\, dP.
\end{equation}
Figure~\ref{fig:absorption} shows this transmission function versus
the ratio of fibre to absorption length, $L_F/\Lambda_m$. A simple
exponential fit, $T \propto \exp\left[-L_F/\Lambda_{\it eff}\right]$,
applied to the points results in an effective attenuation length of
$\Lambda_{\it eff}= 2.4$\,m. This description is sufficient to
parameterise the transmission function for $L_F/\Lambda_m \ge 0.2$, at
lower values the light is attenuated faster. The difference of order
15\% to the meridional attenuation length is attributed to the tail of
the path length distribution.

\subsection{Propagation of Photons}

The analysis of trapped photons is based on the total photon path
length per axial fibre length, $P$, the number of internal reflections
per axial fibre length, $\eta$, and the optical path length between
successive internal reflections, $l_R$, where we follow the
nomenclature of Potter and Kapany. It should be noted that these three
variables are not independent: $P= \eta \times l_R$.

\begin{figure}[htbp]
  \begin{center}
    \epsfig{width= 0.5 \textwidth, file= 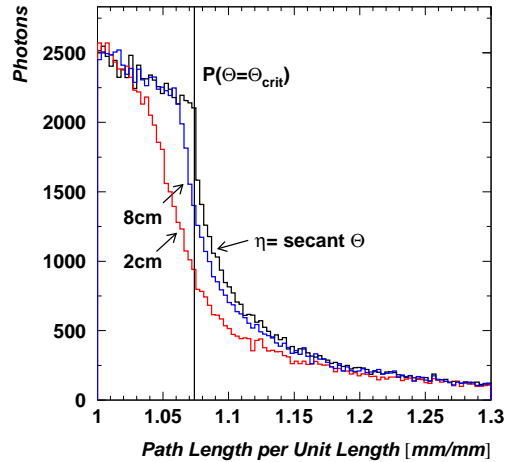}
    \caption{ The distribution of the optical path length, $P(\theta)$, 
	of trapped photons in fibres of radius $\rho=$ 0.6\,mm 
	normalised to the axial length of the fibre. The figure shows 
	$P(\theta)$ for a straight fibre and for two different radii 
	of curvature, $R_{\it curve}=$ 2 and 8\,cm. The vertical line at 
	$P(\theta_{\it crit})=$ 1.074 indicates the upper limit of 
	$P$ in the meridional approximation.}
    \label{fig:pathlength}
  \end{center}
\end{figure}

Figure~\ref{fig:pathlength} shows the distribution of the normalised
optical path length, $P(\theta)$, for photons reaching the exit end of
straight and curved fibres of 0.6\,mm radius. The figure also gives
results for curved fibres of two different radii of curvature. The
distribution of path lengths shorter than the path length for
meridional photons propagating at the critical angle is almost
flat. It can easily be shown that the normalised path length along a
straight fibre is given by the secant of the axial angle and is
independent of other fibre dimensions: $P(\theta)= \sec\theta$. In
case of the curved fibre the normalised path length of the trapped
photons is less than the secant of the axial angle and photons on near
meridional paths are refracted out of the fibre most.

\begin{figure}[htbp]
  \begin{center}
    \epsfig{width= 0.5 \textwidth, file= 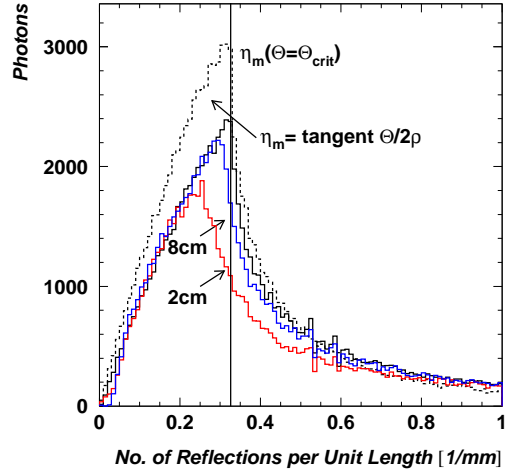}
    \caption{ The distribution of the number of reflections, 
	$\eta(\theta)$, for trapped photons in fibres of radius 
	$\rho=$ 0.6\,mm normalised to the axial length of the fibre. 
	The figure shows $\eta(\theta)$ for a straight fibre and 
	for two different radii of curvature, $R_{\it curve}=$ 2 
	and 8\,cm. The vertical line at $\eta_m (\theta_{\it crit})=$ 
	0.326 indicates its upper limit in the meridional approximation. 
	The dashed line shows the distribution of 
	$\eta_m(\theta)= \tan{\theta}/2\rho$.}
    \label{fig:reflections}
  \end{center}
\end{figure}

The distribution of the normalised number of reflections,
$\eta(\theta)$, for photons reaching the exit end of straight and
curved fibres is shown in figure~\ref{fig:reflections}. Again, the
figure gives results for curved fibres of two different radii of
curvature. The number of reflections a photon experiences scales with
the reciprocal of the fibre radius. In the meridional approximation
the normalised number of reflections is related by simple trigonometry
to the axial angle and the fibre radius: $\eta_m(\theta) =
\tan{\theta}/2\rho$. The distribution of $\eta_m$, based on the
distribution of axial angles for the trapped photons, is represented
by the dashed line. The upper limit, $\eta(\theta_{\it crit})$, is
indicated in the plot by a vertical line. The number of reflections
made by a skew ray, $\eta_s(\theta)$, can be calculated for a given
skew angle: $\eta_s(\theta)= \eta_m(\theta) / \cos{\gamma}$. It is
clear that this number increases significantly if the skew angle
increases. From the distributions it can be seen that in curved fibres
the trapped photons experience fewer reflections on average.

\begin{figure}[htbp]
  \begin{center}
    \subfigure[]{
      \epsfig{width= 0.47 \textwidth, file= 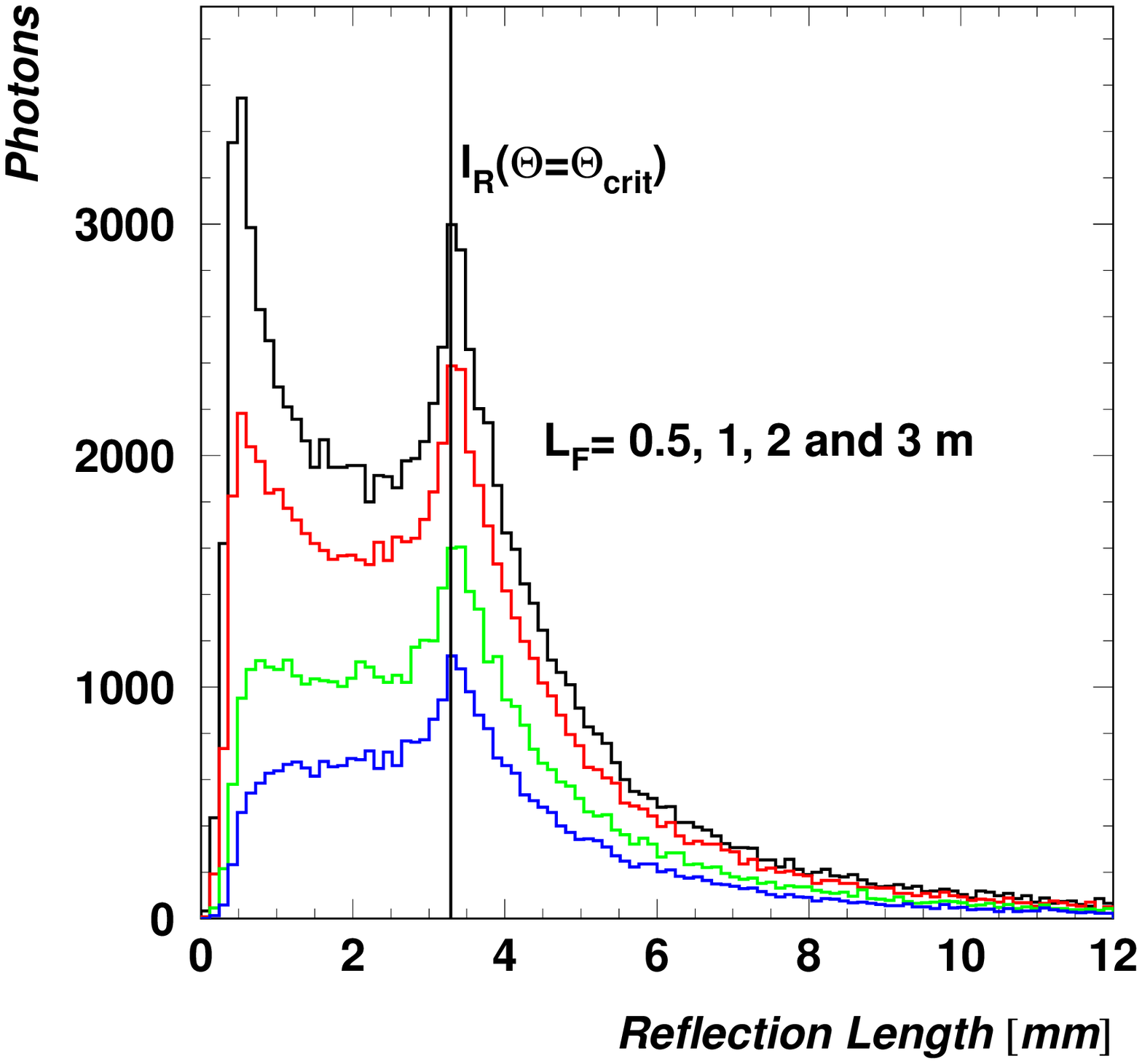} }
    \subfigure[]{
      \epsfig{width= 0.47 \textwidth, file= 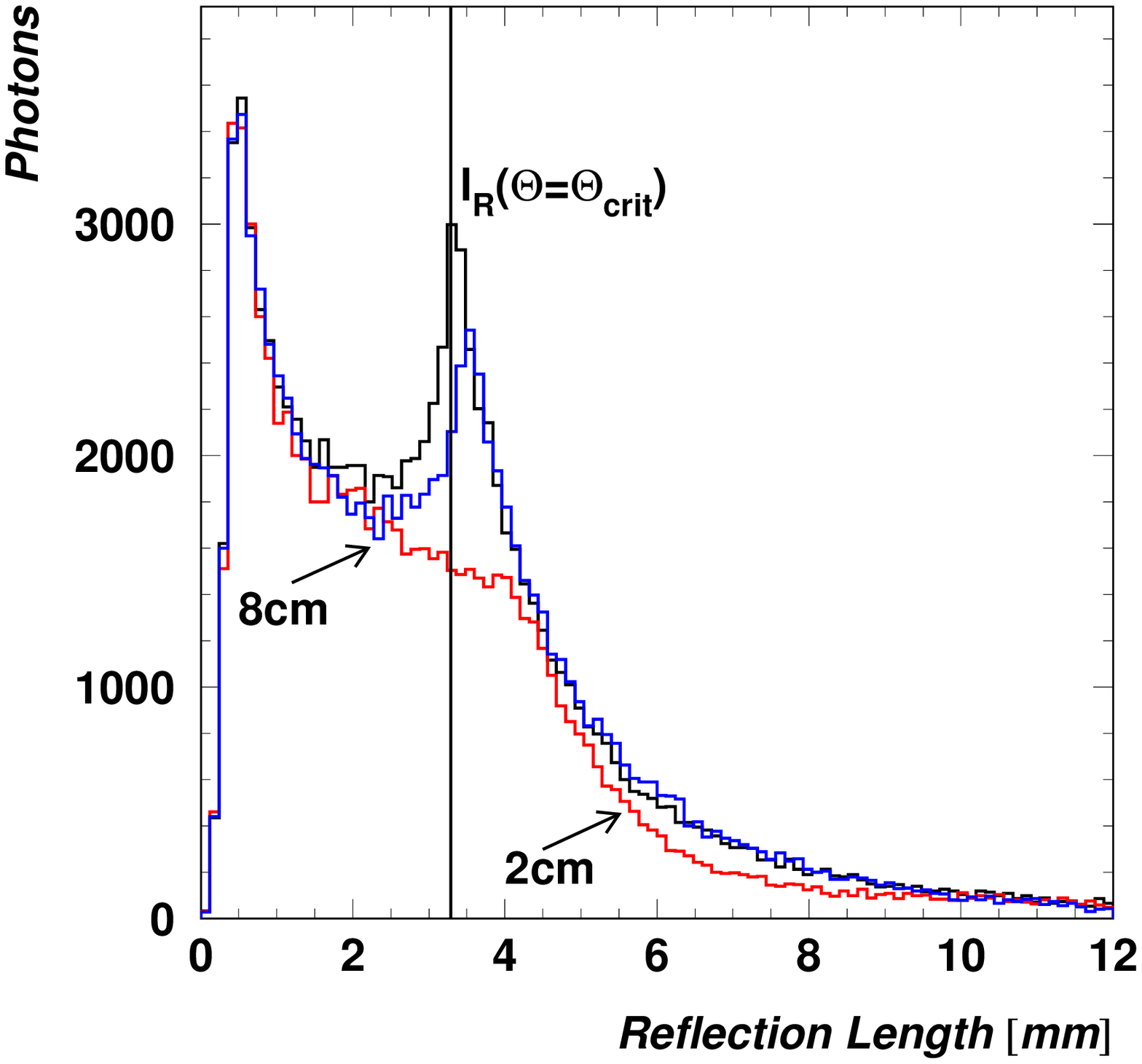} }
    \caption{ The distribution of the reflection length, 
	$l_R(\theta)$, for photons reaching the exit end of 
	fibres of radius $\rho=$ 0.6\,mm. The figure shows 
	$l_R(\theta)$ for straight fibres (a) with three 
	different fibre lengths, $L_F=$ 0.5, 1, 2 and 3\,m and 
	for curved fibres (b) with radii of curvature $R_{\it curve}=$ 
	2 and 8\,cm. The vertical line at $l_R(\theta_{\it crit})=$ 
	3.29\,mm indicates the lower limit of $l_R$ in the meridional 
	approximation.}
    \label{fig:rlambda}
  \end{center}
\end{figure}

Figure~\ref{fig:rlambda} shows the distribution of the reflection
length, $l_R(\theta)$, for photons reaching the exit end of fibres of
radius $\rho= 0.6$\,mm. The reflection length will scale with the
fibre radius. The left figure shows $l_R(\theta)$ for three different
over-all fibre lengths and the attenuation behaviour of the photons is
made apparent by the non-vanishing attenuation parameters used. Short
reflection lengths correspond to long optical path lengths and large
numbers of reflections. Because of the many reflections and the long
total paths traversed, these photons will be attenuated faster than
photons with larger reflection lengths. This reveals the high
attenuation of rays with large skew angles. In the meridional
approximation the reflection length is related to the axial angle by:
$l_R= 2\rho/\cos{\theta}$. In the figure the minimum reflection length
allowed by the critical angle condition is shown by a vertical line at
$l_R(\theta_{\it crit})= 3.29$\,mm. The right figure shows
$l_R(\theta)$ in sharply curved fibres of two different radii of
curvature.  It can be seen that the region of highest attenuation is
close to the reflection length for photons propagating at the critical
angle. On average photons propagate with smaller reflection lengths
along the curved fibre.

In contrast to the analysis of straight fibres an approximation of the
sharply curved fibre by meridional rays is not a very good one, since
only a very small fraction of the light rays have paths lying in the
bending plane. It is clear that when a fibre is curved the path
length, the number of reflections and the reflection length of a
particular ray in the fibre are affected, which is clearly seen in
Figs.~\ref{fig:pathlength},~\ref{fig:reflections} and
\ref{fig:rlambda}, where the over-all fibre length is 50\,cm. The
average optical path length and the average number of reflections in a
fibre curved over a circular arc are less than those for the same ray
in a straight fibre for those photons which remain trapped.

\subsection{Trapping Efficiency and Losses in Sharply Curved Fibres}

One of the most important practical issues in implementing optical
fibres into compact particle detector systems are macro-bending
losses. In general, some design parameters of fibre applications,
especially if the over-all size of the detector system is important,
depend crucially on the minimum permissible radius of curvature. By
using the waveguide analysis a loss formula in terms of the Poynting
vector can be derived~\cite{Marcuse1976}, but studies on bending
losses in single-mode fibres cannot be directly applied to large
diameter fibres. By using ray optics for those fibres the analysis of
the passage of skew rays along a curved fibre becomes highly complex.

\begin{figure}[htbp]
  \begin{center}
    \epsfig{width= 0.6 \textwidth, file= 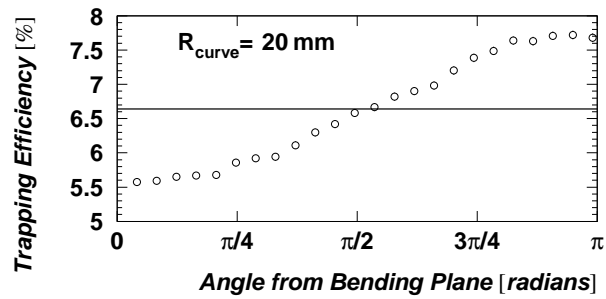}
    \caption{ Trapping efficiencies for photons in a sharply 
	curved fibre with radius of curvature $R_{\it curve}=$ 
	2\,cm and fibre radius $\rho=$ 0.6\,mm. 
	The trapping efficiency is plotted as a function of the 
	angle between the photon path and the bending plane, so 
	that 0\,rad corresponds to photons emitted towards the 
	tensile side of the fibre.}
    \label{fig:trap-bend}
  \end{center}
\end{figure}

The angle of incidence of a light ray at the tensile (outer) side of
the fibre is always smaller than at the compressed side and photons
propagate either by reflections on both sides or in the extreme
meridional case by reflections on the tensile side only. If the fibre
is curved over an arc of constant radius of curvature photons can be
refracted, and will then no longer be trapped, at the very first
reflection point on tensile side. Therefore, the trapping efficiency
for photons entering a curved section of fibre towards the tensile
side is reduced most. Figure~\ref{fig:trap-bend} quantifies the
dependence of the trapping efficiency on the azimuthal angle between
the bending plane and the photon path for a curved fibre with a radius
of curvature $R_{\it curve}=$ 2\,cm.

\begin{figure}[htbp]
  \begin{center}
    \epsfig{width= 0.6 \textwidth, file= 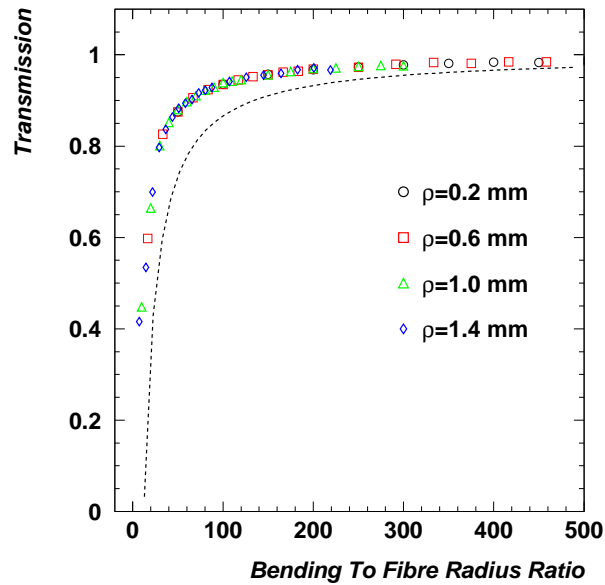}
    \caption{ Characteristics of bending losses in curved fibres.
	The transmission function for fibres curved over a circular arc 
	of 90\,$^\circ$ is plotted as a function of the radius of 
    	the curvature to fibre radius ratio for different fibre radii, 
	$\rho=$ 0.2, 0.6, 1.0 and 1.4\,mm. The dashed line is a simple
	estimate from the meridional approximation.}
    \label{fig:bradius}
  \end{center}
\end{figure}

Figure~\ref{fig:bradius} displays the explicit dependence of the
transmission function for fibres curved over circular arcs of
90$^\circ$ on the radius of curvature to fibre radius ratio for
different fibre radii, $\rho=$ 0.2, 0.6, 1.0 and 1.2\,mm. No further
light attenuation is assumed. Evidently, the number of photons which
are refracted out of a sharply curved fibre increases very rapidly
with decreasing radius of curvature. The losses are dependent only on
the ratio, since no inherent length scale is involved, justifying the
introduction of the curvature to fibre radius ratio as a scaling
variable. The light loss due to bending of the fibre is about 10\% for
a radius of curvature of 65 times the fibre radius.

\begin{figure}[htbp]
  \begin{center}
    \epsfig{width= 0.3 \textwidth, file= 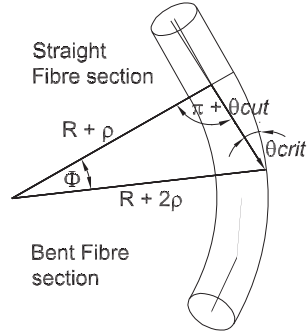}
    \caption{ Extreme geometric condition for the trapping of a 
	photon in a curved fibre with radius $\rho$ and radius 
	of curvature $R_{\it curve}$.}
    \label{fig:sketch}
  \end{center}
\end{figure}

In the meridional approximation in the bending plane a cut-off angle
\begin{equation}
  \cos\theta_{\it cut}= \frac{R + 2 \rho}{R + \rho} 
  \cos\theta_{\it crit} 
  \label{eq:transmission}
\end{equation}
can be introduced so that photons emitted with this axial angle
$\cos\theta_{\it cut}$ are reflected at the tensile side of the fibre
with the critical angle.  Figure~\ref{fig:sketch} shows the passage of
this meridional ray through the fibre. Reference~\cite{Gloge1972}
makes a comparison of the meridional approximation with the waveguide
analysis and justifies its use.  A transmission function can be
estimated from this cut-off angle by assuming that all photons with
axial angles $\theta > \theta_{\it cut}$ are refracted out of the
fibre:
\begin{equation}
  T= 1 - \frac{1}{1 + R_{\it curve}/\rho}\ 
  \frac{\cos\theta_{\it crit}}{1 - \cos\theta_{\it crit}}.
\end{equation} 
This transmission function is shown in figure~\ref{fig:bradius} and it
is obvious that it is overestimating the light losses, because of the
larger axial angles allowed for skew rays.

The first point of reflection for photons entering a curved section of
fibre can be characterised by the bending angle, $\Phi$, through which
the fibre is bent. Close to the critical angle the angular phase space
density is highest. For photons emitted towards the tensile side the
corresponding bending angle is related to the axial angle of the
photon, $\Phi= \theta_{\it crit} - \theta_{\it cut}$. For a fibre
radius $\rho=$ 0.6\,mm and radii of curvature $R_{\it curve}=$ 1, 2,
and 5\,cm the above formula leads to bending angles $\Phi=$ 0.19, 0.08
and 0.03\,rad, respectively.

\begin{figure}[htbp]
  \begin{center}
      \epsfig{width= 0.55 \textwidth, file= 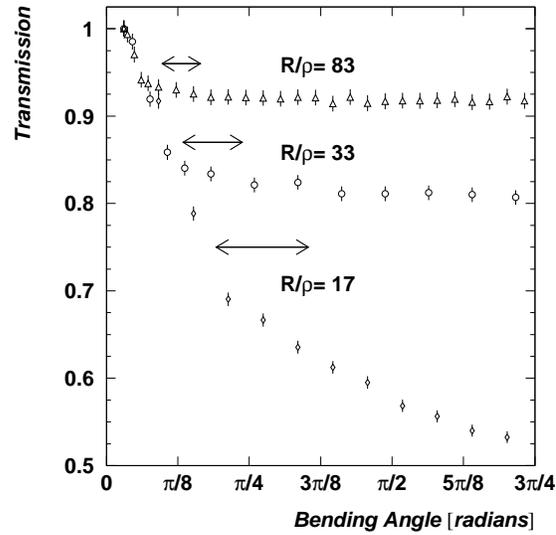}
      \caption{ Simulated transmission function for a curved fibre 
	of radius $\rho=$ 0.6\,mm with three different radii of 
        curvature, $R_{\it curve}=$ 1, 2 and 5\,cm, corresponding to 
	the ratios $R_{\it curve}/\rho=$ 17, 33 and 83, respectively. 
	The ordinate is the fraction of photons reaching the fibre exit 
	end as a function of the bending angle, $\Phi$, and the arrows 
	indicate angles, $\Phi_{\it limit}$, where in the bending plane 
	a change in the transmission function should occur.}
      \label{fig:bending}
  \end{center}
\end{figure}

Photons emitted from the fibre axis towards the compressed side are
not lost at this side, however, they experience at least one
reflection on the tensile side if the bending angle exceeds the limit
$\Phi_{\it limit} = \arccos\left[R_{\it curve}/(R_{\it curve} + 2\,
\rho)\right] \approx \arccos\left[1 - 2\, \rho / R_{\it
curve}\right]$. A change in the transmission function should occur at
bending angles between $\Phi_{\it limit}/2$, where all photons emitted
towards the tensile side have experienced a reflection, and $\Phi_{\it
limit}$, where this is true for all photons. Figure~\ref{fig:bending}
shows the transmission as a function of bending angle, $\Phi$, for the
same fibre conditions as before. Once a sharply curved fibre with a
ratio $R_{\it curve}/\rho > 83$ is bent through angles $\Phi \sim
\pi/8$\,rad light losses do not increase any further. The limiting
angles range from $\Phi_{\it limit}= 1.06$\,rad to $\Phi_{\it limit}=
0.44$\,rad and are indicated in the figure by arrows. At much smaller
ratios $R_{\it curve}/\rho$ the meridional approximation is no longer
valid to describe this initial behaviour.

Contours of phase space distributions for photons refracted out of
sharply curved fibres with radii of curvature $R_{\it curve}=$ 2 and
5\,cm are shown in figure~\ref{fig:phasespace}. The contours
demonstrate that skew rays from a small region close to the
$\cos{\overline{\gamma}}$ curve are getting lost. The smaller the
radius of curvature, the larger the affected phase space region.

\subsection{Light Dispersion}

A pulse of light, consisting of several photons propagating along a
fibre, broadens in time. In active fibres, three effects are
responsible for the time distribution of photons reaching the fibre
exit end. Firstly the decay time of the fluorescent dopants, usually
of the order of a few nanoseconds, secondly the wavelength spectrum of
the emitted light, which leads to different propagation velocities for
different photons in a dispersive medium, and thirdly the fact that
photons on different paths have different transit times to reach the
fibre exit end, known as inter-modal dispersion. The timing resolution
of scintillators are often of paramount importance.

\begin{figure}[htbp]
  \begin{center}
      \epsfig{width= 0.55 \textwidth, file= 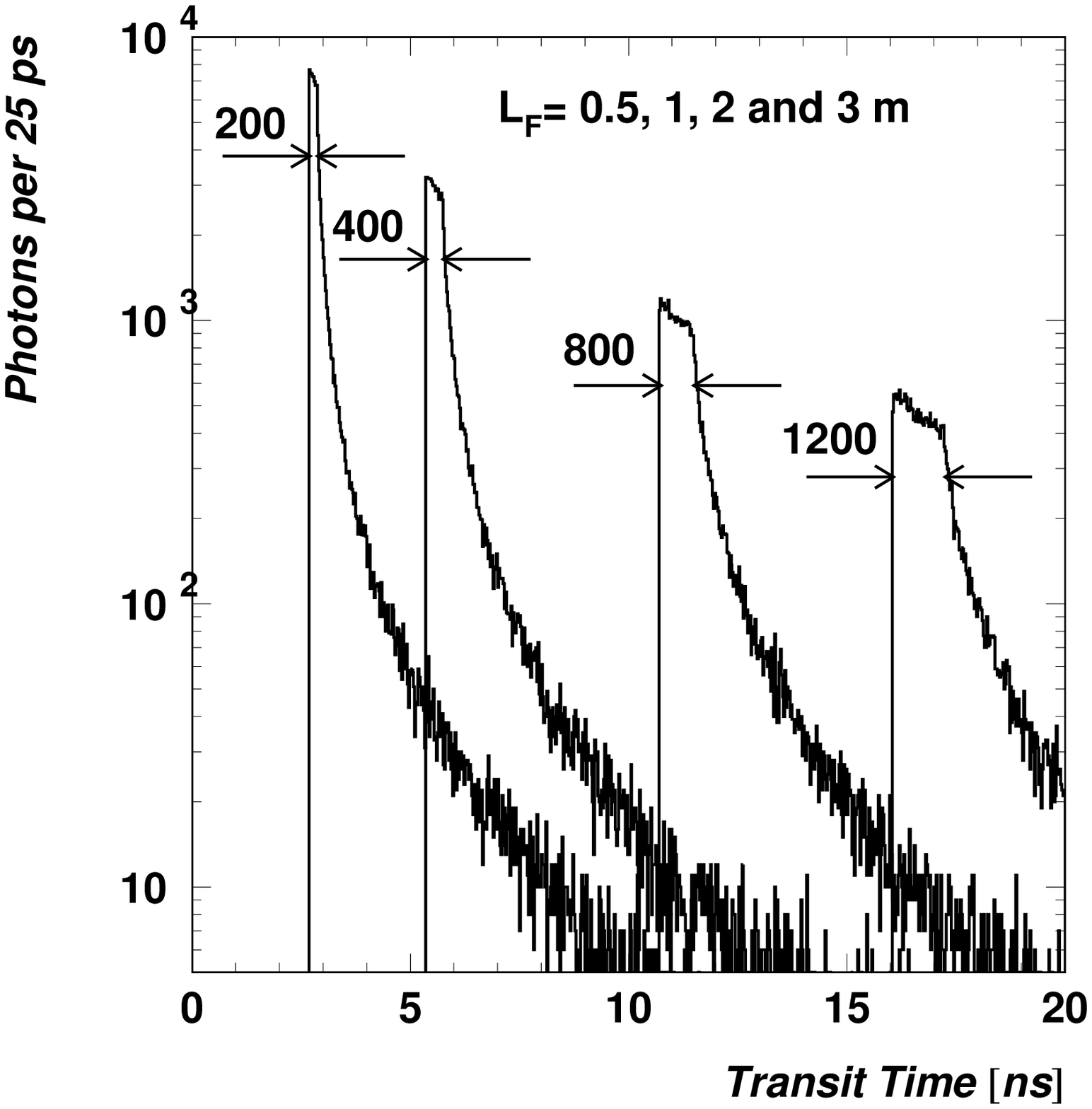}
      \caption{ The distribution of the transit time in nanoseconds 
	for photons reaching the fibre exit end. For the fibre lengths
	$L_F=$ 0.5, 1, 2 and 3\,m the pulse dispersion (FWHM) of the 
	transit time distribution is 200, 400, 775, and 1200\,ps, 
	respectively.}
      \label{fig:timing}
  \end{center}
\end{figure}

The transit time in ray optics is simply given by $\tau=
P(\theta)/c_{\it core}$, where $c_{\it core}$ is the speed of light in
the fibre core. The simulation results on the transit time are shown
in figure~\ref{fig:timing}. The full widths at half maximum (FWHM) of
the pulses in the time spectrum are presented for four different fibre
lengths. The resulting dispersion has to be compared with the time
dispersion in the meridional approximation which is simply the
difference between the shortest transit time $\tau(\theta= 0)$ and the
longest transit time $\tau(\theta= \theta_{\it crit})$: $\Delta \tau=
L_F (\sec{\theta_{\it crit}}-1)/c_{\it core}$, where $L_F$ is the
total axial length of the fibre. The dispersion evaluates for the
different fibre lengths to 197\,ps for 0.5\,m, 393\,ps for 1\,m,
787\,ps for 2\,m and 1181\,ps for 3\,m. Those numbers are in good
agreement with the simulation, although there are tails associated to
the propagation of skew rays. With the attenuation parameters of our
simulation the fraction of photons arriving later than $\tau(\theta=
\theta_{\it crit})$ decreases from 37.9\% for a 0.5\,m fibre to 32\%
for a 3\,m fibre due to the stronger attenuation of the skew rays in
the tail.

\section{Summary}

We have simulated the propagation of photons in straight and curved
optical fibres. The simulations have been used to evaluate the loss of
photons propagating in fibres curved in a circular path in one
plane. The results show that loss of photons due to the curvature of
the fibre is a simple function of radius of curvature to fibre radius
ratio and is $< 10\%$ if the ratio is $> 65$. The simulations also
show that for larger ratios this loss takes place in the initial stage
of the bend ($\Phi \sim \pi/8$) during which a new distribution of
photon angles is established. The photons which survive this initial
loss then propagate without further bending losses.

We have also used the simulation to investigate the dispersion of
transit times of photons propagating in straight fibres. For fibre
lengths between 0.5 and 3\,m we find that approximately two thirds of
the photons arrive within the spread of transit times which would be
expected from the use of the simple meridional ray approximation and
the refractive index of the fibre core. The remainder of the photons
arrive in a tail at later times due to their helical paths in the
fibre. The fraction of photons in the tail of the distribution
decreases only slowly with increasing fibre length and will depend on
the attenuation parameters of the fibre.

We find that when realistic bulk absorption and reflection losses are
included in the simulation for a straight fibre, the overall
transmission can not be described by a simple exponential function of
propagation distance because of the large spread in optical path
lengths between the most meridional and most skew rays.

We anticipate that these results on the magnitude of bending losses
will be of use for the design of particle detectors incorporating
sharply curved active fibres.

\ack This research was supported by the UK Particle Physics and Astronomy
Research Council (PPARC).


\clearpage
\newpage


%
%


%
%


%
%


%


\section*{References}
\begin{thebibliography}{14}

\bibitem{Leutz1995}
Leutz H, Scintillating fibres, {\it Nucl. Instr. and Meth. in Phys. Res.} 
	{\bf A364} (1995) 422--448.

\bibitem{Snitzer1961}
Snitzer E, Cylindrical dielectric waveguide modes, {\it J. Opt. Soc. Am.}
	{\bf 51~(5)} (1961) 491--498.

\bibitem{Kapany1963}
Kapany N~S, Burke J~J and Shaw C~C, Fiber optics. {X}. {E}vanescent boundary
  	wave propagation, {\it J. Opt. Soc. Am.} {\bf 53~(8)} (1963) 929--935.

\bibitem{Kapany1957}
Kapany N~S, Fiber optics. {I}. {O}ptical properties of certain dielectric
  	cylinders, {\it J. Opt. Soc. Am.} {\bf 47~(5)} (1957) 413--422.

\bibitem{Kapany1967}
Kapany N~S, Fibre Optics: Principles and Applications, Academic Press, London
	and New York, 1967.

\bibitem{Allan1973}
Allan W~B, Fibre Optics: Theory and Practice, Optical Physics and
	Engineering, Plenum Press, London and New York, 1973.

\bibitem{Potter1961}
Potter R~J, Transmission properties of optical fibers, {\it J. Opt. Soc. Am.}
	{\bf 51~(10)} (1961) 1079--1089.

\bibitem{Kapany1961}
Kapany N~S and Capellaro D~F, Fiber optics. {VII}. {I}mage transfer from
  	{L}ambertian emitters, {\it J. Opt. Soc. Am.}{\bf  51~(1)} 
	(1961) 23--31, (appendix: Geometrical optics of straight circular 
	dielectric cylinder).

\bibitem{Potter1963}
Potter R~J, Donath E and Tynan R, Light-collecting properties of a perfect
  	circular optical fiber, {\it J. Opt. Soc. Am.} {\bf 53~(2)} (1963)
	256--260.

\bibitem{Recipes1992}
Press W~H, Teukolsky S~A, Vetterling W~T and Flannery B~P, Numerical
  	Recipes in Fortran77: The Art of Scientific Computing, 2nd Edition, 
	Vol. 1 of Fortran Numerical Recipes, Cambridge University Press, 1992.

\bibitem{Johnson1994}
Johnson K~F, Achieving the theoretical maximum light yield in scintillating
  	fibres through non-uniform doping, {\it Nucl. Instr. and Meth. in 
	Phys. Res.}{\bf  A344} (1994) 432--434.

\bibitem{Ambrosio1991}
D'Ambrosio C, Leutz H and Taufer M, Reflection losses in polysterene fibres,
  	{\it Nucl. Instr. and Meth. in Phys. Res.} {\bf A306} (1991) 549--556.

\bibitem{Marcuse1976}
	Marcuse D, Curvature loss formula for optical fibers, {\it J. Opt. 
	Soc. Am.} {\bf 66~(3)} (1976) 216--220.

\bibitem{Gloge1972}
Gloge D, Bending loss in multimode fibers with graded and ungraded core index,
  	{\it Appl. Opt.} {\bf 11~(11)} (1972) 2506--2513.

\end{thebibliography}
\end{document}